\documentclass[a4paper,fleqn,usenatbib]{mnras}
\usepackage{Times}
\usepackage[T1]{fontenc}
\usepackage{ae,aecompl}

\usepackage{graphicx}	
\usepackage{amsmath}	
\usepackage{amssymb}	

\title[Differential evolution of the Ly$\alpha$ LF at $z=5.7-6.6$]{The Ly$\alpha$ luminosity function at $\bf z=5.7-6.6$ and the steep drop of the faint end: implications for reionization}

\author[Santos, Sobral \& Matthee]{S\'{e}rgio Santos$^{1,2,3}$\thanks{E-mail: ssantos@oal.ul.pt}, David Sobral$^{3,4}$ \& Jorryt Matthee$^{4}$  \\ 
$^{1}$ Instituto de Astrof\'{\i}sica e Ci\^{e}ncias do Espa\c{c}o, Universidade de Lisboa, OAL, Tapada da Ajuda, PT1349-018 Lisboa, Portugal \\
$^{2}$ Departamento de F\'{i}sica, Faculdade de Ci\^{e}ncias, Universidade de Lisboa, Edif\'{i}cio C8, Campo Grande, PT1749-016 Lisbon, Portugal \\
$^{3}$ Department of Physics, Lancaster University, Lancaster, LA1 4YB, UK \\
$^{4}$ Leiden Observatory, Leiden University, P.O.\ Box 9513, NL-2300 RA Leiden, The Netherlands}

\pubyear{2016}

\begin{document}
\label{firstpage}
\pagerange{\pageref{firstpage}--\pageref{lastpage}}
\maketitle

\begin{abstract}
We present new results from the widest narrow band survey search for Ly$\alpha$ emitters at $z=5.7$, just after reionization. We survey a total of 7\,deg$^2$ spread over the COSMOS, UDS and SA22 fields. We find over 11,000 line emitters, out of which 514 are robust Ly$\alpha$ candidates at $z=5.7$ within a volume of $6.3\times10^6$ Mpc$^3$. Our Ly$\alpha$ emitters span a wide range in Ly$\alpha$ luminosities, from faint to bright (L$_{\rm Ly\alpha}\sim10^{42.5-44}$\,erg\,s$^{-1}$) and rest-frame equivalent widths (EW$_0\sim25-1000$\,\AA) in a single, homogeneous data-set. By combining all our fields we find that the faint end slope of the $z=5.7$ Ly$\alpha$ luminosity function is very steep, with $\alpha=-2.3^{+0.4}_{-0.3}$. We also present an updated $z=6.6$ Ly$\alpha$ luminosity function, based on comparable volumes and obtained with the same methods, which we directly compare with that at $z=5.7$. We find a significant decline of the number density of faint Ly$\alpha$ emitters from $z=5.7$ to $z=6.6$ (by $0.5\pm0.1$\,dex), but no evolution at the bright end/no evolution in $L^*$. Faint Ly$\alpha$ emitters at $z=6.6$ show much more extended haloes than those at $z=5.7$, suggesting that neutral Hydrogen plays an important role, increasing the scattering and leading to observations missing faint Ly$\alpha$ emission within the epoch of reionization. All together, our results suggest that we are observing patchy reionization which happens first around the brightest Ly$\alpha$ emitters, allowing the number densities of those sources to remain unaffected by the increase of neutral Hydrogen fraction from $z\sim5$ to $z\sim7$.
\end{abstract}


\section{Introduction} \label{sec:introduction}

During the past two decades, considerable progress has been made in understanding the distant/early Universe \citep[see reviews by e.g.][]{Robertson2010,Dunlop2012,Madau2014}. Currently, the samples of $z>6$ candidates are mostly composed by rest-frame ultra-violet (UV) selected galaxies obtained from extremely deep surveys with the Hubble Space Telescope \citep[e.g.][]{Bouwens2015,Finkelstein2015}. However, spectroscopy and multi-wavelength follow-up \citep[e.g. with ALMA;][]{Ouchi2013,Watson2015,Capak2015,Maiolino2015} of these sources still remains very limited as most candidates are too faint for a detailed analysis with current instrumentation \citep[see also][]{Dunlop2016}. Alternatively, emission lines can be used to search for high-redshift galaxies to directly select galaxies by their brightest features, including several rest-frame optical and UV lines \citep[e.g.][]{Ouchi2008,Sobral2013,Khostovan2015,Khostovan2016}, allowing for efficient follow-up strategies.

The Lyman-$\alpha$ (Ly$\alpha$) emission line (rest-frame 1215.67\,\AA) is emitted by both young star-forming galaxies and active galactic nuclei/quasars, being intrinsically the strongest emission line in the rest-frame optical to UV \citep[e.g.][]{Partridge1967,Pritchet1994}. As Ly$\alpha$ is redshifted into optical wavelengths (it can be observed from the ground at $z\approx2-7$), many other strong lines are redshifted out of even the near-infrared \citep[see e.g.][]{Ly2007,Ly2011, Hayes2010, Sobral2013}, making Ly$\alpha$ one of the only available means of spectroscopic confirmation, along with other weaker high ionisation UV lines \citep[e.g.][]{Sobral2015,Stark2016}.

Several approaches have been used to find and study Ly$\alpha$ emitters, including blind spectroscopy \citep[e.g.][]{Martin2004,Stark2007,Rauch2008,Sawicki2008,Bayliss2010,Cassata2011}, narrow band surveys \citep[e.g.][]{Cowie1998, Rhoads2000, Rhoads2003, Malhotra2004, Taniguchi2005, Shimasaku2006, Westra2006, Iye2006, Nilsson2007, Murayama2007, Ouchi2008, Ouchi2010, Sobral2009, Hu2010, Kashikawa2011, Shibuya2012, Konno2014, Matthee2014, Matthee2015} and Integral Field Unit (IFU) observations \citep[e.g.][]{Blanc2011,Adams2011,vanBreukelen2005,Bacon2015,Karman2015}. Blind spectroscopy and IFU surveys can be very efficient at probing ultra-low luminosity sources at a variety of redshifts, but the current small volumes probed make them unable to reach even $L^*$ sources, as the rarer (brighter) sources have number densities several times smaller that these studies can reach. Wide narrow band surveys can be very competitive at efficiently probing large volumes at specific look-back times, and can be used to study a much larger luminosity range. For example, one MUSE pointing \citep[e.g.][]{Bacon2015} probes a volume of $\sim10^3$\,Mpc$^3$ for $z\sim3-6$, while one Subaru Suprime-Cam pointing with a typical narrow band filter probes a volume of $\sim10^5$ Mpc$^3$ (Hyper Suprime-Cam covers a volume $\sim$ 7 times larger per pointing). Typically, narrow band surveys have targeted a maximum of $\sim$ 1\,deg$^2$ areas, corresponding to maximum volumes of $\sim10^6$ Mpc$^3$ \citep[e.g.][]{Ouchi2008, Ouchi2010}, but the next generation of surveys are now starting to probe much larger volumes \citep[e.g.][]{Matthee2015,Hu2016}.

Due to its resonant nature, Ly$\alpha$ photons are easily scattered by neutral hydrogen \cite[and also easily absorbed by dust; e.g.][]{Hayes2011}. As a consequence, the observability of Ly$\alpha$ can in principle be used as a probe of the neutral state of the inter-galactic medium (IGM) during the epoch of reionization \citep[e.g.][]{Fontana2010, Caruana2012, Schenker2012, Ono2012, Caruana2014, Dijkstra2014,Pentericci2014,Schmidt2016}. However, in order to interpret Ly$\alpha$ observations (such as the distribution of equivalent widths, the fraction of UV selected galaxies with strong Ly$\alpha$, or the evolution of the number density of Ly$\alpha$ emitters) as consequences of reionization, one needs to accurately understand the contribution from potentially varying intrinsic inter-stellar medium (ISM) properties such as the Ly$\alpha$ escape fraction \citep[c.f.][]{Matthee2016}, overdensities of galaxies (e.g. \citealt{Castellano2016}) or selection biases in UV selected galaxy samples \citep[c.f.][]{Oesch2015,Zitrin2015,Stark2016}. Therefore, it is important to have a clear understanding of Ly$\alpha$ with only little influence from the IGM at $z\approx6$, when reionization is close to complete and the fraction of neutral hydrogen becomes extremely low \citep{Fan2006, Becker2015}.

Previous studies found that the Ly$\alpha$ luminosity function (LF) seems to have little evolution at $z\sim3-6$ \citep[e.g.][]{Ouchi2008}. In contrast, the UV LF of Lyman-break galaxies (LBGs) strongly decreases for higher redshifts \citep[e.g.][]{Bouwens2015,Finkelstein2015}. This difference in evolution is likely explained by an evolving escape fraction of Ly$\alpha$ photons, likely due to a lower dust content, younger stellar populations, lower metallicities and/or a combination of related phenomena. This is consistent with the observation that the fraction of LBGs with strong Ly$\alpha$ emission increases up to $z=6$ \citep[e.g.][]{Stark2010,Cassata2015}. At $z>6$ the number density of faint Ly$\alpha$ emitters (LAEs) is found to decline with redshift \citep{Ouchi2010,Konno2014}, likely due to reionization not being fully completed. However, by using the largest Ly$\alpha$ survey at $z\sim7$ ($\sim5$\,deg$^2)$, \cite{Matthee2015} show that the strong decrease/evolution in the number density of LAEs happens pre-dominantly at relative faint Ly$\alpha$ luminosities, while the bright end (with luminosities L$_{\rm Ly\alpha}>10^{43}$\,erg\,s$^{-1}$) may not evolve at all. \cite{Matthee2015} finds that bright LAEs at $z=6.6$ are much more common than previously thought, with spectroscopic confirmation presented in \cite{Sobral2015}, and with independent studies finding consistent results \citep[see e.g.][]{Hu2016}. However, one strong limitation in interpreting the potential evolution from $z=6.6$ to $z=5.7$ is the lack of comparably large $\sim5-10$\,deg$^2$, multiple field surveys that can both trace a large enough number of bright sources and overcome cosmic variance.

In this work, we present the largest Ly$\alpha$ narrow band survey at $z=5.7$, covering a total of $\sim7$\,deg$^2$ ($\sim10^7$\,Mpc$^3$). Previous studies have never probed beyond 2\,deg$^2$ \citep[e.g.][]{Murayama2007,Ouchi2008,Hu2010}, and have mostly focused on specific, single fields. Here we take advantage of previous data and add further $\sim4$\,deg$^2$ of unexplored data. We also re-analyse the $z=6.6$ luminosity function presented in \cite{Matthee2015}.

We structure this paper as follows: Section \ref{sec:observations} presents the observations and data reduction. Section \ref{sec:criteria} explains the selection of line emitters and Ly$\alpha$ emitters at $z=5.7$. In Section \ref{sec:LF} we present the method and procedures adopted to construct the $z=5.7$ and $z=6.6$ Ly$\alpha$ LFs. We present our results in Section \ref{results}, including a comparison with previous surveys. Section \ref{sec:discussion} discusses the results in the context of predicted effects from reionization. Finally Section \ref{sec:conclusions} presents the conclusions of this paper.

Throughout this work we use a $\Lambda$CDM cosmology with H$_0 = 70$\,km\,s$^{-1}$\,Mpc$^{-1}$, $\Omega _M = 0.3$ and $\Omega _\Lambda = 0.7$. All magnitudes in this paper are presented in the AB system. At $z=5.7$, 1$''$ corresponds to 5.9 kpc.

\section{Observations and Data Reduction} \label{sec:observations}

\subsection{Observations}

We have reduced and analyzed raw archival NB816 data in the COSMOS, UDS and SA22 fields. We use these three fields as they are completely independent (preventing any possible bias from probing the same region of the sky) and far enough from the galactic plane (avoiding bright foreground stars and dust). Additionally, the available deep multi-wavelength coverage (including optical and near infra-red) allows a robust selection of candidates and identification of any lower redshift interlopers.

%
%
\begin{table*}
\caption{Our NB816 data in the COSMOS, UDS and SA22 fields. The SA22 field was separated into two sub-fields, deep and wide, according to its significantly different NB816 depth. R.A. and Dec. are the central coordinates of the fields. FWHM is the average value for the seeing and is similar across our entire coverage. The NB816 depth is the 2$\sigma$ depth measured in 2$''$ apertures. Note that the quoted area already takes into account the removed/masked regions which are not used in this paper.}
\begin{center}
\begin{tabular}{lccccc}
\hline
\bf Field &  \bf R.A. & \bf Dec. & {\bf Area} & \bf FWHM  & \bf NB816 depth\\
 & (J2000) &   (J2000)   &  (deg$^2$)  &  ($''$) & (2$\sigma$, 2$''$)\\
\hline
COSMOS &10 00 00 &+02 10 00&2.00&0.7 & 26.2 \\
UDS &02 18 00 & $-$05 00 00 &0.85& 0.7 &26.1 \\
SA22-deep &22 18 00 &+00 20 00 &0.55&0.7& 26.1\\
SA22-wide &22 15 00 &+00 50 00 &3.60 &0.5& 25.0 \\
\hline
\end{tabular}
\label{tab:observations}
\end{center}
\end{table*} 

\begin{figure}
\centering
\includegraphics[width=8.3cm]{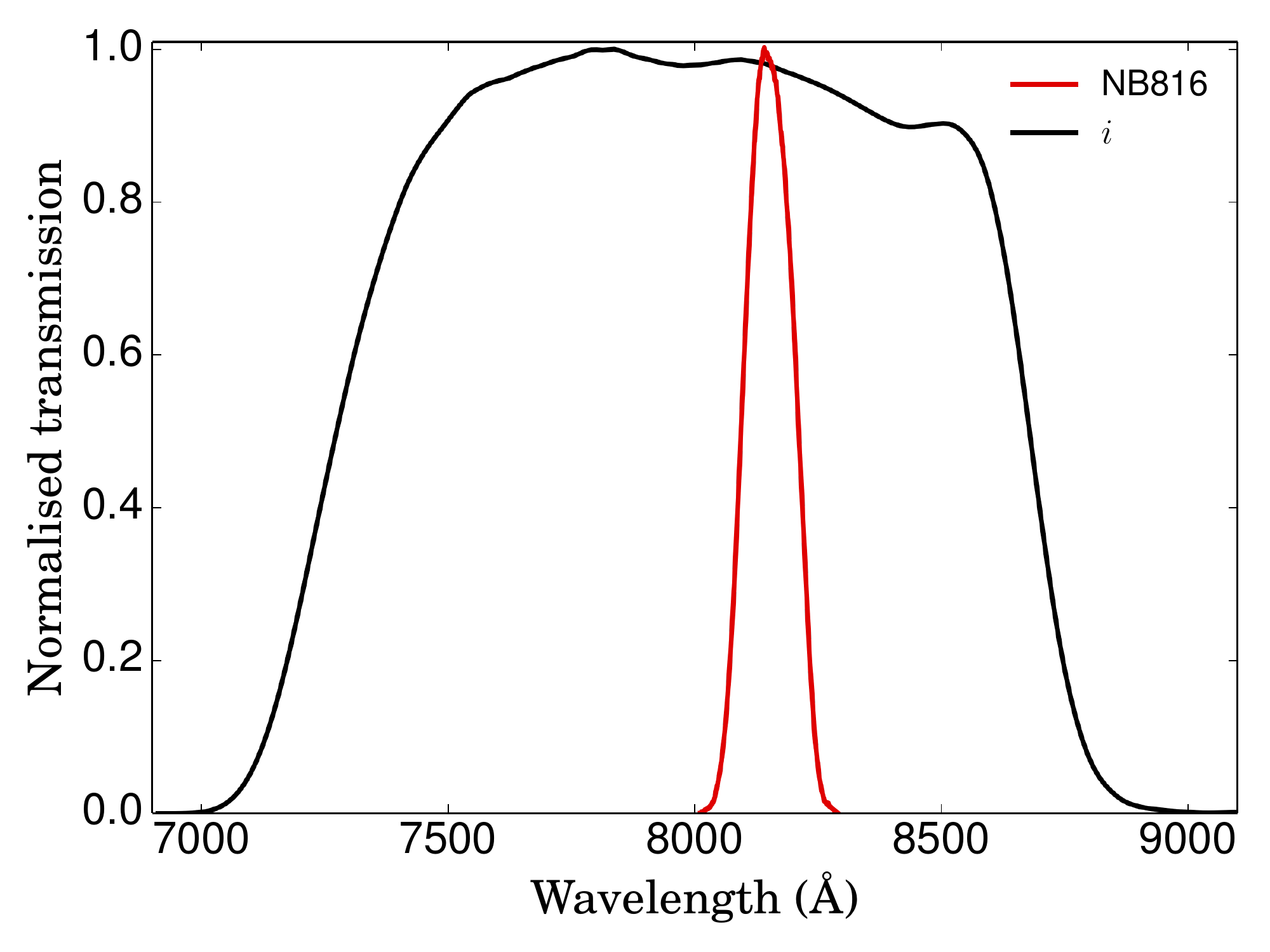}
\caption{Normalised filter profiles of the NB816 and the $i$ band filters used in this study. We note that the shown $i$ band is for Subaru's Suprime-Cam after the upgrade to red sensitive CCDs, such that its peak is slightly shifted towards the red compared to the CFHT MegaCam $i$ band used for SA22. Our NB correction in Section \ref{sec:nbcorrection} takes this into account. NB816 is contained slightly red from the center of $i$. The NB816 filter is located in a wavelength region free of strong atmospheric OH lines.}
 \label{fig:filters}
\end{figure}

The NB816 filter has a central wavelength of 8150\,\AA\, and a full width at half maximum (FWHM) of 120\,\AA. NB816 is contained within the red wing of the broad band filter $i$ (see Figure \ref{fig:filters}). All NB816 data were collected with the Suprime-Cam instrument from the Subaru Telescope \citep{Miyazaki2002}. Suprime-Cam has ten 2048x4096 CCDs arranged in a 5$\times$2 pattern, with a corresponding field of view of $\sim0.25$\,deg$^2$. We use a total of 30 of these pointings. Suprime-cam images have a pixel scale of 0.20$''$\,pix$^{-1}$.

We retrieved all publicly available raw NB816 data for the UDS and SA22 fields from the SMOKA Archive\footnote{http://smoka.nao.ac.jp/}. Fully reduced COSMOS NB816 images (original PSF) were retrieved from the COSMOS Archive\footnote{http://irsa.ipac.caltech.edu/data/COSMOS/} \citep{Taniguchi2007,Capak2007}.

We split SA22 data into two different sub-fields (SA22-deep and SA22-wide), which differ in depth by $\approx1$\,mag and in area by a factor of $\approx6.6$. SA22-wide contains the largest area (larger than COSMOS and UDS combined). Narrow band observations are summarized in Table \ref{tab:observations}.

Previous studies have separately used NB816 data in COSMOS \citep[][]{Murayama2007}, UDS/SXDF \citep[][]{Ouchi2008} and SA22-deep \citep[$\sim0.4$ deg$^2$;][]{Hu2010}. We note that while we explore new data and provide the largest survey of its kind, we are able to reproduce individual results from the literature using our own analysis. A comparison between our findings and previous studies is presented in Section \ref{sec:LF}.

%
%

\subsection{Data reduction}

We used the Subaru data reduction pipelines \citep[{\sc sfred} and {\sc sfred2};][]{Ouchi2004} to reduce the NB816 data. The data reduction follows the same procedure as detailed in e.g. \cite{Matthee2015} and we refer the reader to that study for more details. Briefly, the reduction steps include: overscan and bias subtraction, flat fielding, point spread function homogenisation, sky background subtraction and bad pixel masking. After these steps, we apply an astrometric calibration using {\sc scamp} \citep{Bertin2006} to correct astrometric distortions. The software matches our images with the 2MASS catalog in the $J$ band \citep{Skrutskie2006} and fits polynomial functions that correct for any distortions along the CCD.

We calibrated the photometry in our data by matching relatively bright, un-saturated stars and galaxies to public catalogues for COSMOS \citep{Laigle2016}, UDS \citep{Cirasuolo2007} and SA22 \citep[][]{Sobral2013,Sobral2015,Matthee2014} using {\sc stilts} \citep{Taylor2006}. NB816 images were calibrated using $i$ band photometry, but a further correction to this calibration was applied in Section \ref{sec:nbcorrection}. Co-added stacks of NB816 exposures were obtained using the {\sc swarp} software \citep{Bertin2002}.

%
%
\begin{figure*}
\includegraphics[width=13.5cm]{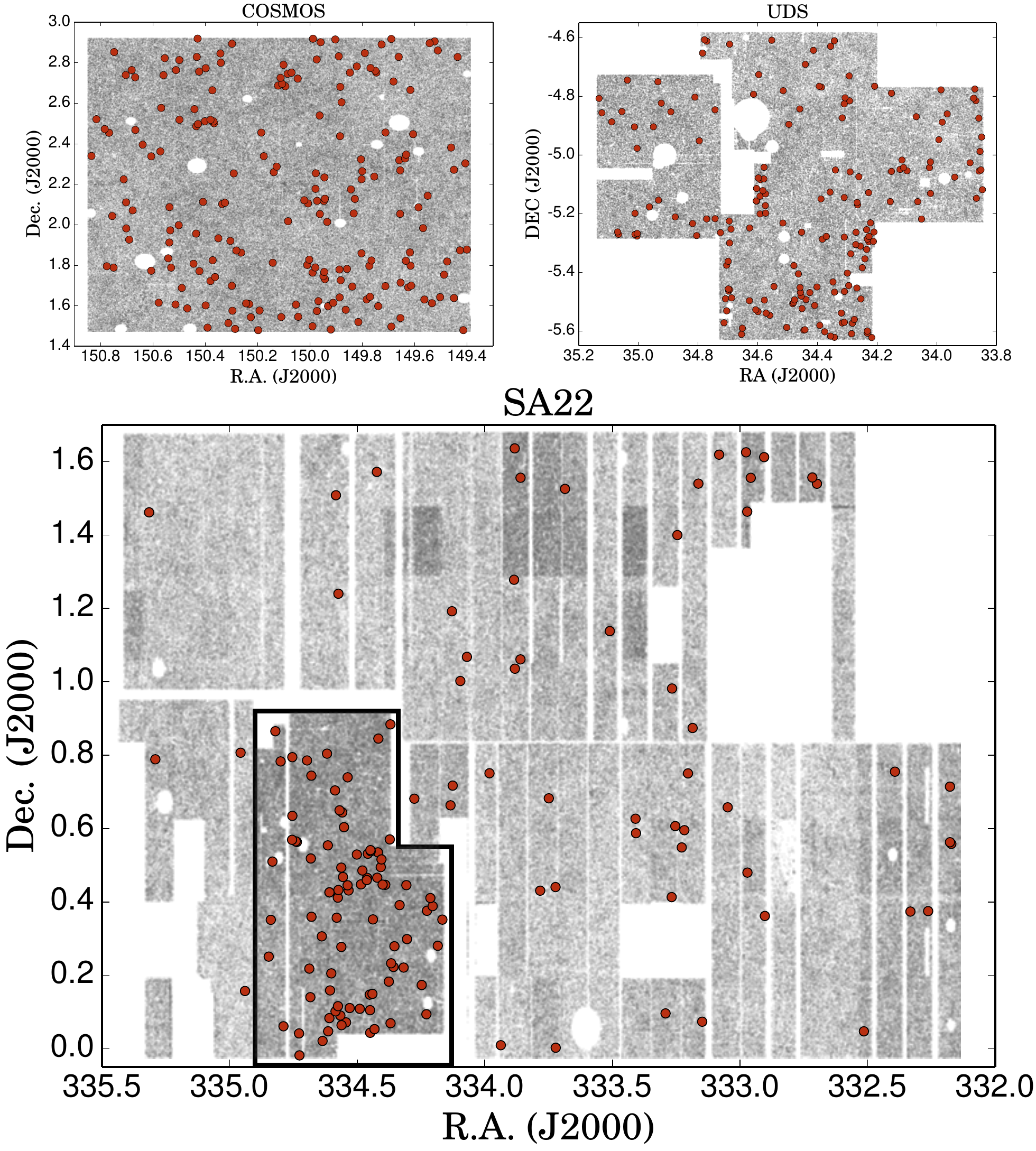}
\caption{The spatial distribution of sources in the COSMOS, UDS and SA22 fields. Grey dots indicate all detections and red circles identify our $z=5.7$ Ly$\alpha$ emitter candidates. A black line contour identifies SA22-deep, the deepest region in the SA22 field. The figure also highlights the regions masked due to bright stars, bad regions and/or low S/N due to dither strategy. It can be seen that UDS, COSMOS and SA22-deep are the deepest regions with a high concentration of sources and candidate LAEs.}
\label{fig:radec}
\end{figure*}

We masked low quality regions, bright haloes around bright stars, diffraction patterns and low S/N regions due to dithering strategy (particularly important in SA22-wide). We also removed regions with low quality or absent $i$ band coverage, regardless of the quality of the narrow band.

We note that our masking is very conservative and, consequently, a relatively large area is removed from our study (hundreds of arcmin$^{2}$), but that is still only a small fraction of our total area. After masking low quality regions, our NB816 coverage contains a total area of 7\,deg$^2$ (Figure \ref{fig:radec}), corresponding to a comoving volume of $6.3\times10^6$\,Mpc$^3$ at $z=5.7$. All areas and volumes used and mentioned in this paper take into account these masks, unless stated otherwise.

Finally, we measure the depth of our images using randomly placed 2$''$ apertures. In each image, we place 200,000 empty apertures in random positions. The average results per field are given in Table \ref{tab:observations}.

%
%
\begin{table*}
\caption{Multi-wavelength depths (2$\sigma$; measured in 2$''$ empty apertures) for the available broad-band filters across all three fields.}\label{tab:depth}
\begin{center}
\begin{tabular}{ccc}
\hline
\bf Field & \bf Broad band filters & \bf Broad band depth (2$\sigma$, 2$''$) \\
\hline
COSMOS & $BVgrizYJHK$ & 27.6, 27.0, 27.1, 27.0, 26.6, 25.7, 25.3, 24.6, 25.0, 24.7 \\
UDS & $BVrizJK$ & 27.5, 27.2, 27.0, 26.8, 27.0, 25.3, 24.8 \\
SA22 & $ugrizJK$ & 26.2, 26.5, 25.9, 25.6, 24.5, 24.3, 23.8  \\
\hline
\end{tabular}
\end{center}
\end{table*}

%
%
\begin{figure*}
\includegraphics[width=16cm]{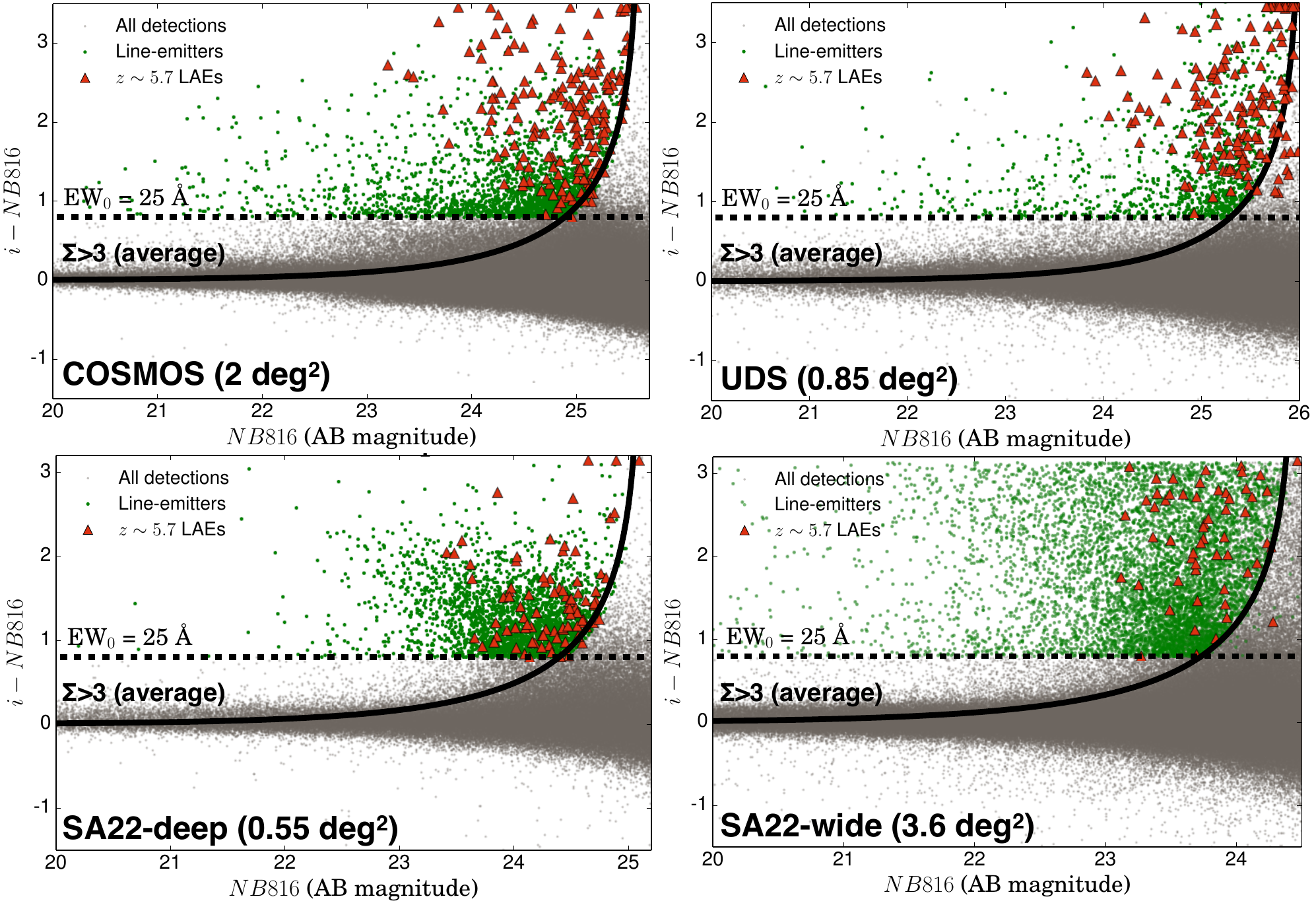}
\caption{Narrow band excess diagram for COSMOS, UDS and SA22. We plot narrow band excess (i broad band magnitude minus NB816 magnitude) versus narrow band NB816 magnitude. Grey points represent all detections after masking, removal of sources with non-physical narrow band and cosmic rays. Green points represent line-emitters, obtained by applying the EW and $\Sigma$ cuts described in Section \ref{sec:criteria}. For visual reference, we collapsed the points with no i detection in the top region of the plots. The $\Sigma$ line shown in this figure is the median value from small sub-fields which we created inside each field.}
\label{fig:excess}
\end{figure*}

\subsection{Multi-wavelength imaging} \label{subsec:mw}

A large collection of multi-wavelength data are publicly available for our entire coverage. For the COSMOS field, we use optical \textit{BVgriz} data taken with the Subaru/SuprimeCam \citep{Taniguchi2007,Capak2007}, retrieved from the COSMOS Archive and NIR \textit{YHJK} data from UltraVISTA DR2 \citep{McCracken2012}, taken with VISTA/VIRCAM. For the UDS field, we use optical \textit{BVriz} data from SXDF \citep{Furusawa2008} and NIR \textit{JHK} data from UKIDSS \citep[][]{Lawrence2007}. For the SA22 field we use optical \textit{ugriz} data from CFHTLS\footnote{http://www.cfht.hawaii.edu/Science/CFHTLS/}, taken with the MegaCam \citep{Boulade2003} and NIR \textit{JK} data from UKIDSS DXS \citep{Warren2007}, taken with UKIRT/WFCAM \citep{Casali2007}. All data which were not taken with the Subaru/SuprimeCam were degraded to a pixel scale of 0.20$''$\,pix$^{-1}$ using {\sc swarp}. A summary of the available filters for each field and their photometric depth is shown in Table \ref{tab:depth}.

\subsection{NB816 catalogue}

The extraction of sources was conducted using {\sc SExtractor} \citep{Bertin1996} in dual extraction mode, using NB816 as the detection image.

\subsubsection{Narrow band magnitude correction} \label{sec:nbcorrection}

The NB816 filter is located slightly to the red of the Subaru Suprime-cam $i$ filter (with red sensitive CCDs) with a separation of $\approx180$\,\AA\, between the center of the two filters. Calibrating the narrow band magnitude directly to the $i$ band may result in an offset in the magnitudes, particularly for sources with strong colours. We correct the narrow band magnitudes by summing a small correction factor which is estimated from the color of the two adjacent broad bands, $i$ and $z$. To compute this correction, we use sources with $i$, $z$ and NB816 magnitudes between 19 and 24 (not saturated and with high enough S/N). The correction has the following expression:

\begin{equation}
NB_{corrected}=NB+0.4\times(i-z),
\end{equation}
where $NB$, $i$ and $z$ are the 2$''$ magnitudes in the respective bands and $NB_{corrected}$ is the corrected NB816 magnitude. We apply this correction to sources with $i$ and $z$ detections. For the remaining sources, we apply a median correction of $+0.20$. As a result of this correction, there is less scatter in the excess diagram (Figure \ref{fig:excess}). The correction also corrects for the fact that the CFHT MegaCam $i$ band is slightly bluer than Suprime-cam's $i$ band, because this slightly different $i$ band will result in slightly different $i-z$ colours.

Our narrow band correction is an alternative to the correction applied in \citet{Murayama2007} who used a corrected broad band obtained from an $iz$ interpolation. Our narrow band correction corresponds to a $BB_{corrected}=0.6i+0.4z$ which is fully consistent with their interpolation.

\subsubsection{Removal of sources with non-physical narrow band detection} \label{sec:unreasonable}

The wavelengths covered by NB816 are contained inside the $i$ band coverage. This means that sources with NB816 detection should be detected in $i$ as long as the $i$ image is deep enough. For each source we compute the expected $i$ magnitude if it only had emission inside NB816. If the measured $i$ magnitude of a source is fainter than this value and the depth of the $i$ image is sufficient to detect it, we remove it from our sample.
This step mainly removes variable sources (such as supernovae and moving sources) and spurious sources that are detected only in the narrow band images and sources with boosted narrow band emission from e.g. diffraction patterns.

\subsubsection{Cosmic ray removal} \label{sec:cosmicray}

Cosmic rays may become artefacts in images. This problem can be avoided through stacking of several frames. However, in our shallower SA22-wide data, the small number of frames causes a less efficient removal of such artefacts during stacking. We created an automated procedure to identify and remove cosmic rays from our sample.

For each source detected in the NB816 imaging we measure the standard deviation in boxes of $5\times5$ pixels around each source. Cosmic rays can be easily identified by their high standard deviation, several times higher than any real source. We apply a cautious cut to make sure we do not lose any real sources. Since we were cautious with this step, we also visually inspect all the final LAE candidates to identify any cosmic ray that was not excluded.

\section{Selecting NB816 line emitters} \label{sec:criteria}

For the selection of line-emitters, we apply similar criteria to e.g. \cite{Sobral2013} and \cite{Matthee2015}, relying on two parameters: equivalent width (EW) and Sigma ($\Sigma$).
The equivalent width is the ratio between the flux of an emission line and the continuum flux. It can be expressed as:

\begin{equation}
EW_{obs} = \Delta\lambda_{NB}\frac{f_{NB}-f_{BB}}{f_{BB}-f_{NB}(\Delta\lambda_{NB}/\Delta\lambda_{BB})},
\end{equation}
where $\Delta\lambda_{NB}$ and $\Delta\lambda_{BB}$ are the FWHM of the narrow band and broad band filters ($\Delta\lambda_{NB816}= $120 \AA; $\Delta\lambda_{i}= $1349 \AA) and $f_{NB}$ and $f_{BB}$ are the flux densities measured in the two filters.

The second parameter, Sigma \citep[$\Sigma$, e.g.][]{Bunker1995}, is used to assure that the excess of the NB816 relative to the broad-band is significantly above the noise. It can be written as \citep[][]{Sobral2013}:

\begin{equation}
\Sigma = \frac{1-10^{-0.4(BB-NB)}}{10^{-0.4(ZP-NB)}\sqrt{rms_{BB}^{2}+rms_{NB}^{2}}}
\end{equation}
where BB and NB are the broad band and the corrected narrow band magnitudes (in this case, NB816 and $i$), ZP is the zero-point of the image (set to 30) and rms is the root-mean-square of the background of the respective image.

To select our sample of line-emitters, we apply the following selection criteria:

\begin{itemize}
\item $i-NB816>0.8$
\item $\Sigma>3$
\end{itemize}

The narrow band excess criteria $i-NB816>0.8$ corresponds to a rest-frame EW of 25\,\AA\, for a $z=5.7$ LAE. This cut is similar to the one used by \citet{Hu2010} and \citet{Matthee2015} for $z=6.6$ but slightly lower than e.g. \citet{Ouchi2008} ($i-NB816>1.2$) and \citet{Taniguchi2005} ($i-NB816>1$).

We present the narrow band excess diagram in Figure \ref{fig:excess}, highlighting our sample of line-emitters. With our selection criteria we identify over 11,000 candidate line emitters.

\subsection{Photometric and spectroscopic redshifts} \label{subsec:photo}

\begin{figure}
  \centering
  \includegraphics[width=0.45\textwidth]{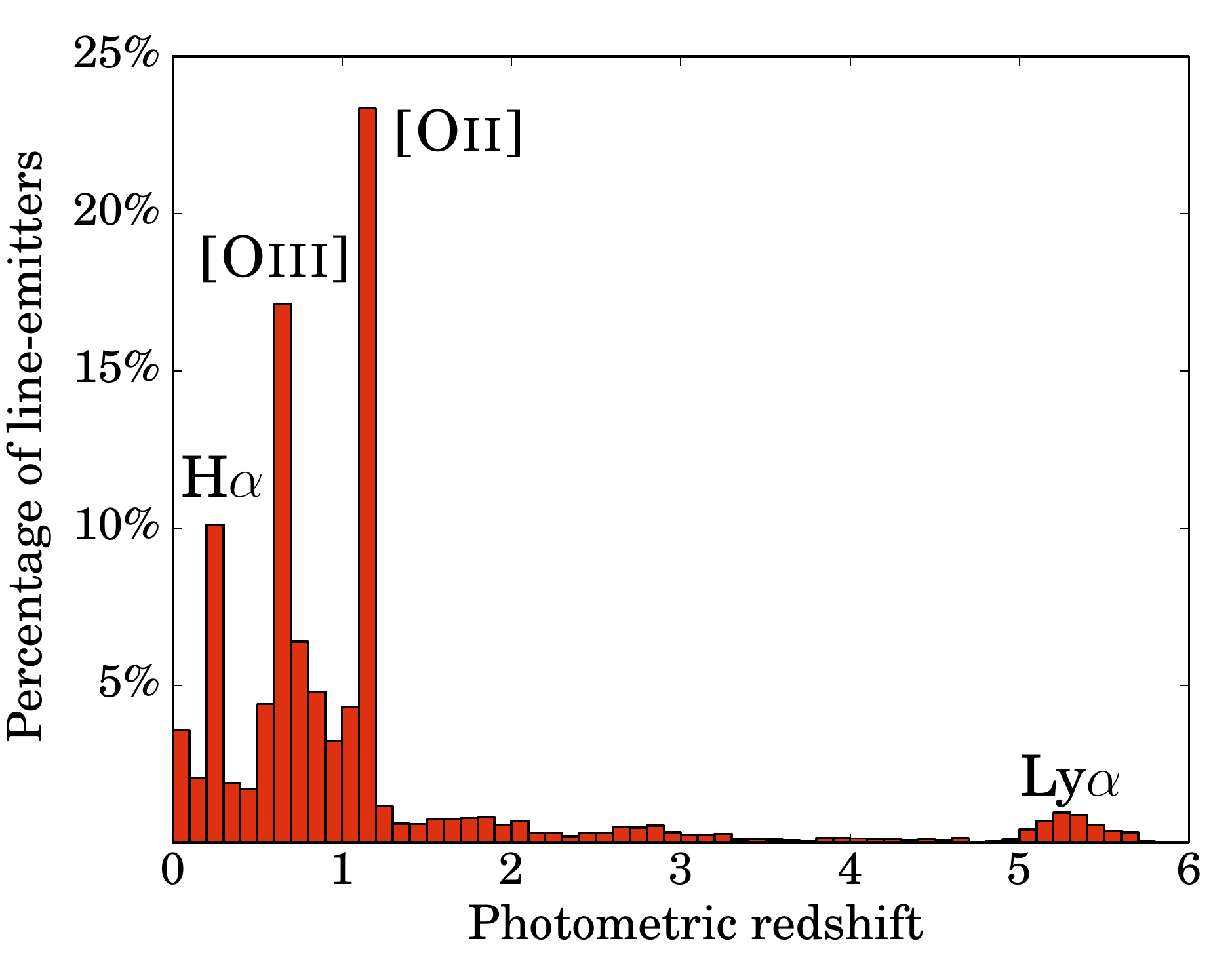}
  \caption{Distribution of photometric redshifts of line-emitters selected in COSMOS, UDS and SA22 by using a simple selection criteria of $i-NB186>0.25$ and $\Sigma>3$. The peaks are consistent with line emission at specific wavelengths. Annotations indicate the redshifts where we expect major emission lines (H$\alpha$ at $z\sim0.2$, [O{\sc iii}] at $\sim0.6$, [O{\sc ii}] at $z\sim1.2$ and Ly$\alpha$ at $z=5.7$).}
 \label{fig:photo}
\end{figure}

In order to explore the nature of the line-emitters, we have used accurate photometric redshifts and a large compilation of spectroscopic redshifts: \cite{Laigle2016} for COSMOS, \cite{Cirasuolo2007} for UDS and a combination of \cite{Kim2015}, \cite{Matthee2014} and \citet{Sobral2015} for SA22. We retrieve $\sim5000$ emitters with either available photometric or spectroscopic redshift. Figure \ref{fig:photo} presents the distribution of photometric redshifts of our sample of line emitters. Even though our high EW cut is tuned to select Ly$\alpha$ emitters at $z=5.7$, our initial sample of line emitters reveals a range of strong line emitters. The peaks in the photometric redshifts are consistent with H$\alpha$ at $z\sim0.2$, {\sc [Oiii]} at $\sim0.6$, {\sc [Oii]} at $z\sim1.2$ and Ly$\alpha$ at $z=5.7$. From our spectroscopic redshift, we find a total of 46 Ly$\alpha$ emitters at $z=5.7$.

As expected, our sample is dominated by lower redshift line-emitters, mostly composed by sources up to $z\sim1.2$. In order to isolate LAEs at z$=5.7$ from our sample we require additional selection criteria, which we will explore in Section \ref{LAEcriteria}.

%
%

\subsection{Selection of LAEs at $\bf z=5.7$} \label{LAEcriteria}

In order to select Ly$\alpha$ emitters and remove low redshift interlopers, we use the Lyman break technique and identify the break at rest frame 912\,\AA, blue-ward of the Lyman limit (although, in practice, at $z=5.7$, radiation blue-ward of Ly$\alpha$ is almost fully absorbed by the Ly$\alpha$ forest; e.g. \citealt{Madau1995}). LAEs at $z=5.7$ should have no strong detection in optical wavelengths below the $i$ band. A weak $r$ band detection is possible if the IGM is relatively transparent (and there are few Ly$\alpha$ forest lines). To summarise, we apply the following criteria, similar to \cite{Ouchi2008}:

\begin{equation} \label{eq:noopticalBV}
B>B_{2\sigma} \wedge V>V_{2\sigma} \wedge [ r>r_{2\sigma} \vee (r<r_{2\sigma} \wedge r-i>1.0)]
\end{equation}
where $B$, $V$, $r$ and $i$ are the 2$''$ magnitudes in the respective bands and the $_{2\sigma}$ subscript indicates the 2$\sigma$ depth for the images of the respective bands (see Table \ref{tab:depth}). As there are no available $BV$ data over the full SA22, we apply a small variation of Equation \ref{eq:noopticalBV} where use $ug$ instead:
\begin{equation} \label{eq:noopticalug}
u>u_{2\sigma} \wedge g>g_{2\sigma} \wedge [ r>r_{2\sigma} \vee (r<r_{2\sigma} \wedge r-i>1.0)]
\end{equation}
where $u$, $g$ are the 2$''$ magnitudes in the respective bands.
This criteria ensures we select sources with no detection in the $BVug$ bands but can have some detection in $r$ as long as there is a strong $i-r$ color break.
 
In extreme cases, $z\sim1$ line emitters with a strong Balmer-break could mimic the Lyman-break that we detect. Fortunately, those sources can be identified by their red colors. Similar to \citet{Matthee2015} we reject sources which have significant red colors in the observed NIR bands. Thus, we consider sources with $J-K>0.5$ to be interlopers. This additional NIR criterion is most important in SA22, where the optical data are relatively shallow.

In order to ensure that our candidates are real detections and not spurious sources, we visually inspect each one of the remaining candidates. We first inspect sources in the narrow band images and reject any fake detections (usually originated by e.g. diffraction patterns from bright sources which were not completely masked). We also visually check that each source does not have an optical detection blue-ward of the Lyman-break. To do so, we create an optical stack using the available optical bands for each field (\textit{BVg} for COSMOS, \textit{BV} for UDS and \textit{ug} for SA22), which significantly increases the depth of our images. 

To summarise, we select line-emitters as Ly$\alpha$ at $z=5.7$ if:

\begin{itemize}

\item They have no optical detection blue-ward of the Lyman-break (Equation \ref{eq:noopticalBV} or \ref{eq:noopticalug}).

\item They satisfy $J-K<0.5$, if detected in the NIR.

\item They pass visual inspection, which includes both reality of NB excess (and checking for variability and/or moving sources) and no detection in optical bands.

\end{itemize}

\subsection{Comparison with other samples of Ly$\alpha$ emitters at $\bf z=5.7$} \label{comparison}

We compare our sample of LAEs with the spectroscopically confirmed sources at $z=5.7$ provided by \cite{Ouchi2008} (UDS), \cite{Hu2010} (SA22-deep) and \cite{Mallery2012} (COSMOS). We find that we recover 46 spectroscopically confirmed sources from previous studies which are above our conservative $\Sigma$ detection threshold (other studies typically only apply an EW cut) and that are not in our conservative masked regions.

\subsection{Final sample of Ly$\alpha$ emitters at $\bf z=5.7$} \label{Final_sample}

Across the COSMOS, UDS and SA22 fields we identify a total of 514 $z=5.7$ LAE candidates (currently 46 are spectroscopically confirmed), spanning a range of Ly$\alpha$ luminosities of $10^{42.5}-10^{44}$\,erg\,s$^{-1}$. We will explore the properties of these sources in the following sections. Table \ref{tab:ncandidates} shows a summary of the number of sources after each selection criterion. The spatial distribution of the LAEs in all fields can be seen in Figure \ref{fig:radec}.

%
%
\begin{table}
\caption{Number of candidates after each selection step. The visual inspections step includes individually checking each source first in both the narrow band NB816 and the broad band $i$ images and then for no detection in the deep optical stacks ($BV$ for UDS, $BVg$ for COSMOS and $ug$ for SA22). Note that due to the shallower broad band data in SA22, a large amount of sources passed the initial filtering, but are rejected with the much deeper $ug$ stacks and our visual checks.}
 \begin{center}
  \label{tab:ncandidates}
  \begin{tabular}{lr}
  \hline
  \bf COSMOS & \bf \# sources\\
  \hline
  $\Sigma>3$, EW$_0>25$\,\AA & 2576\\
  No optical detection & 396\\
  After visual inspections & 192\\
  \hline
  \bf UDS\\
  \hline
  $\Sigma>3$, EW$_0>25$ &  981\\
  No optical detection & 239\\
  After visual inspections & 178\\
  \hline
  \bf SA22-wide\\
  \hline
  $\Sigma>3$, EW$_0>25$ & 4692\\
  No optical detection & 1264\\
  After visual inspections & 56\\
  \hline
  \bf SA22-deep\\
  \hline
  $\Sigma>3$, EW$_0>25$ &  2803\\
  No optical detection & 541\\
  After visual inspections & 88\\
  \hline
  \bf Total Ly$\alpha$ $\bf z=5.7$ ($\bf z_{\rm \bf spec}$ confirmed) & 514 (46)\\
  \hline
\end{tabular}
 \end{center}
\end{table}

\section{Computing the Ly$\alpha$ LF} \label{sec:LF}

\subsection{Completeness correction} \label{sec:compcor}

Faint sources and sources with weak emission lines may be missed by our selection criteria, causing the measured number density of sources to be underestimated. To estimate the line-flux completeness we follow \cite{Sobral2013}, adapted for Ly$\alpha$ studies by \cite{Matthee2015}: we construct a sample of high-redshift non line-emitters selected through a simple color break selection ($r-i>1.5$) and add non-emitters with photometric or spectroscopic redshift higher than 4. Using these sources, in steps of increasing line-flux, we artificially increase their NB816 and $i$ band fluxes and then apply our selection criteria on these simulated sources. By determining the fraction that we retrieve as a function of added line-flux, we obtain a completeness estimation for each luminosity bin, which we apply to each bin in our LF. A higher completeness correction is measured for the fainter sources as they are much easier to be missed. The line-flux completeness per luminosity bin for each field is presented in Table \ref{tab:completeness}. The completeness corrected number counts in the different observed fields as a function of their Ly$\alpha$ luminosity are shown in Figure \ref{fig:lfindividual} and in Table \ref{tab:bins_combined}.

\subsection{Filter profile correction} \label{sec:filtercor}

The narrow band filter transmission NB816 has a gaussian distribution with a lower transmission in the wings (Figure \ref{fig:filters}). Sources which have a redshift in the borders of the filter will only be observed at a fraction of their Ly$\alpha$ luminosity \citep[see e.g.][]{Hu2010}. It is necessary to apply a correction factor that compensates the fact that the filter is not top-hat, otherwise, the number densities of bright LAEs will be systematically underestimated. We apply a correction similar to \cite{Matthee2015}. We use the Schechter fit from our data to generate the Ly$\alpha$ luminosity of 1 million sources at a random redshift between $z=5.65$ and $z=5.75$ (corresponding to the edges of NB816). For each luminosity bin, the correction factor is determined from the detection ratio of these fake sources retrieved with the two different filter profiles. The effect of the filter profile correction of our LF is shown in Figure \ref{fig:filtercorrect}. The correction is higher for the brightest bins as these LAEs will likely be observed at a fraction of their luminosity due to the filter not being top-hat.

\subsection{Aperture corrections} \label{aper_correction}

Due to instrumental/observational effects (e.g. seeing/PSF) and mostly due to Ly$\alpha$ photons easily scattering within haloes, Ly$\alpha$ flux can be significantly extended \citep[e.g.][]{Momose2014,Wisotzki2016,Matthee2016,Borisova2016}. The 2$''$ apertures we use are $3-4\times$ the PSF, and thus for point-like sources we do not expect aperture corrections to be important, but if sources are physically extended, $2''$ apertures may lead to missing flux. We investigate this by comparing the NB816 fluxes measured in 2$''$ with those measured with {\sc mag-auto} and study any necessary correction as a function of observed 2$''$ flux. We find little to no dependence up to at least the highest fluxes, and derive a median correction of +0.02 in Ly$\alpha$ luminosity, which we apply (see further discussion in Section \ref{sec:aperture}).

\subsection{Interloper correction} \label{interlopers}

While in COSMOS and UDS the available broad-band data allows to clearly identify and remove interlopers/lower redshift line emitters, in SA22 this is not necessarily the case, particularly for the sources with the faintest continuum. In order to mitigate this, we use our combined COSMOS and UDS with full information, but study the dataset assuming the depths of broad-band imaging were the same as SA22-deep and SA22-wide. We find that, as expected, the contamination is higher (10\% higher) for SA22-like data-sets. We therefore correct all our luminosity bins in SA22 for this expected extra contamination.

\subsection{Obtaining a comparison LF at $\bf z=6.6$} \label{recomp_z66}

In order to compare our results at $z=5.7$, we explore the results and sample presented by \cite{Matthee2015} and apply any necessary corrections/modifications to derive a new, updated $z=6.6$ LF. We use the same methods for completeness and filter profile corrections. We compute the errors per bin by not only taking into account the Poissonian errors, but also by considering systematic errors due to the completeness and filter profile corrections. Furthermore, following our selection criteria, we also carefully check for any variable sources and/or moving sources which can contaminate the bright end. \cite{Matthee2015} applied a statistical correction for these potential contaminants, but we chose to investigate sources one by one, following what we do at $z=5.7$. We note that such statistical correction works very well for COSMOS and UDS, but is a slight underestimation for SA22, as the number of moving sources in SA22 is significantly higher. Nonetheless, we find that none of the results from \citet{Matthee2016}, which are based on spectroscopic follow-up \citep{Sobral2015}, have significantly changed: luminous LAEs (L$_{\rm Ly\alpha}>10^{43.5}$\,erg\,s$^{-1}$) at $z=6.6$ are more common ($\gtrsim30$ times) than previously measured by smaller area studies \citep[e.g.][]{Ouchi2010}. We note that we also apply an aperture correction to the $z=6.6$ LF of +0.11, unchanged from \cite{Matthee2015}.

\section{Results} \label{results}

\subsection{The $\bf z=5.7$ Ly$\alpha$ luminosity function}

\subsubsection{Field to field variations}

We group our LAEs in luminosity bins according to their Ly$\alpha$ luminosity. The observed number density in each bin is corrected for its corresponding line-flux completeness correction. We only include sources from sub-fields with a completeness higher than 25\%. The number density for each luminosity bin is calculated by multiplying the number of counts by the completeness factor, divided by the probed volume and bin width. The errors are Poissonian, but we add 30\% of the completeness correction in quadrature to obtain the final error per bin.

%
%
\begin{figure}
  \centering
  \includegraphics[width=8.3cm]{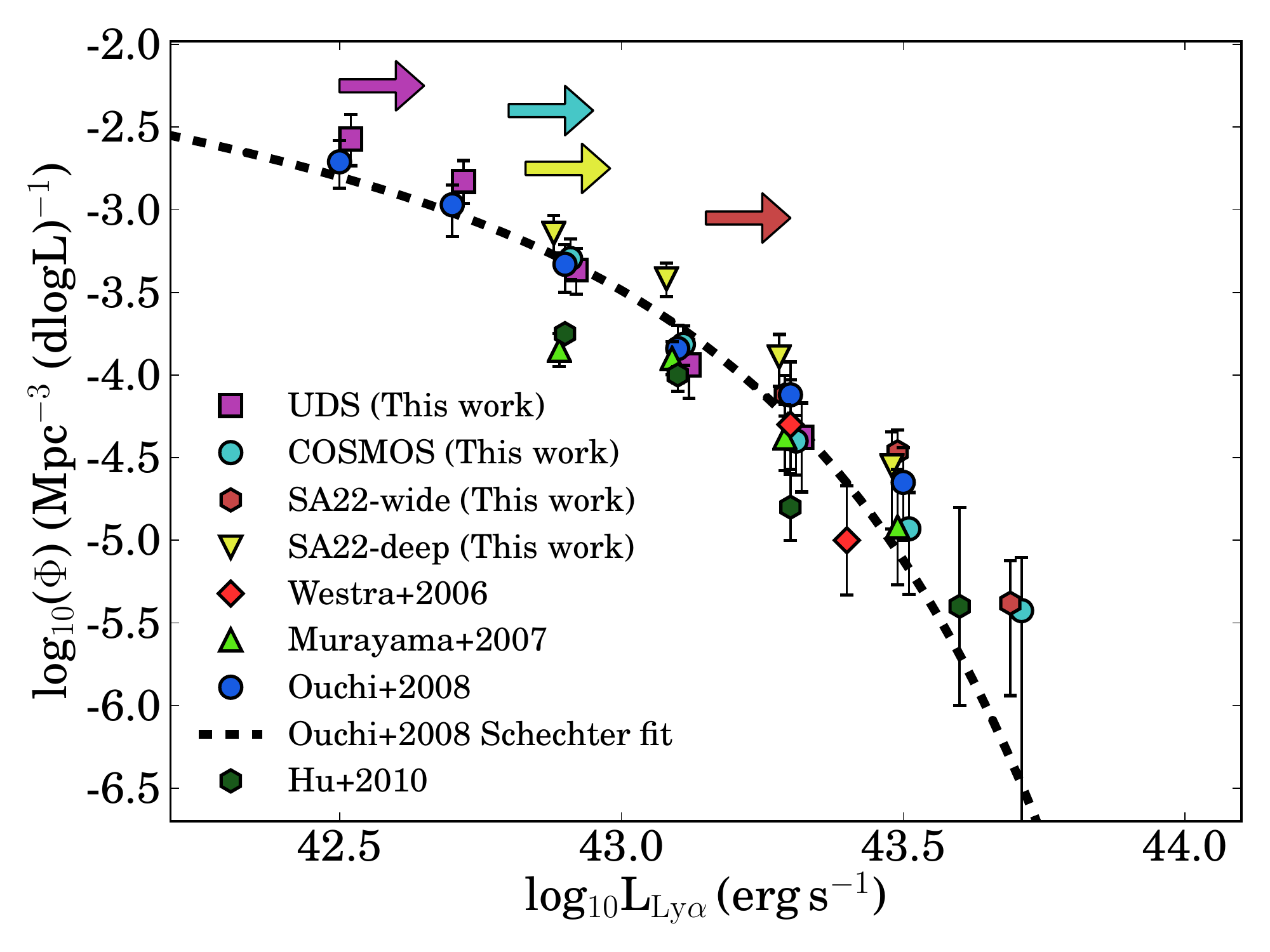}
  \caption{The Ly$\alpha$ luminosity function at $z=5.7$ based on different fields. For visual reference, a small offset in the luminosities ($\pm0.02$\,dex) was used to minimize overlapping of points in the figure. The arrows indicate the luminosity bins for which each field has an average completeness higher than 25\%. We find significant field to field variations of $\pm0.4$\,dex in number densities, consistent with results from e.g. \citet{Ouchi2008}. We also compare our results per field with previous studies, finding them to be consistent with \citet{Murayama2007} and \citet{Ouchi2008}. However, by probing larger, multiple volumes we overcome cosmic variance.}
  \label{fig:lfindividual}
\end{figure}

In Figure \ref{fig:lfindividual} we show the $z=5.7$ Ly$\alpha$ luminosity computed per field. We find that there is significant scatter, of the order of $\pm0.4$\,dex in the number densities, at least for the range of luminosities where we can compare results from all our fields. It may well be that such scatter is reduced for fainter sources, but our sample does not allow us to constrain that as we can only investigate that with a single field (UDS) -- see \cite{Ouchi2008}. Our results per field are also presented in Table \ref{tab:bins}. Our results highlight the importance of probing multiple fields and caution the over-interpretation of single field ``over" or ``under" densities, either in the context of reionization or of structure formation.

%
%
\begin{figure*}
\centering
\includegraphics[width=13cm]{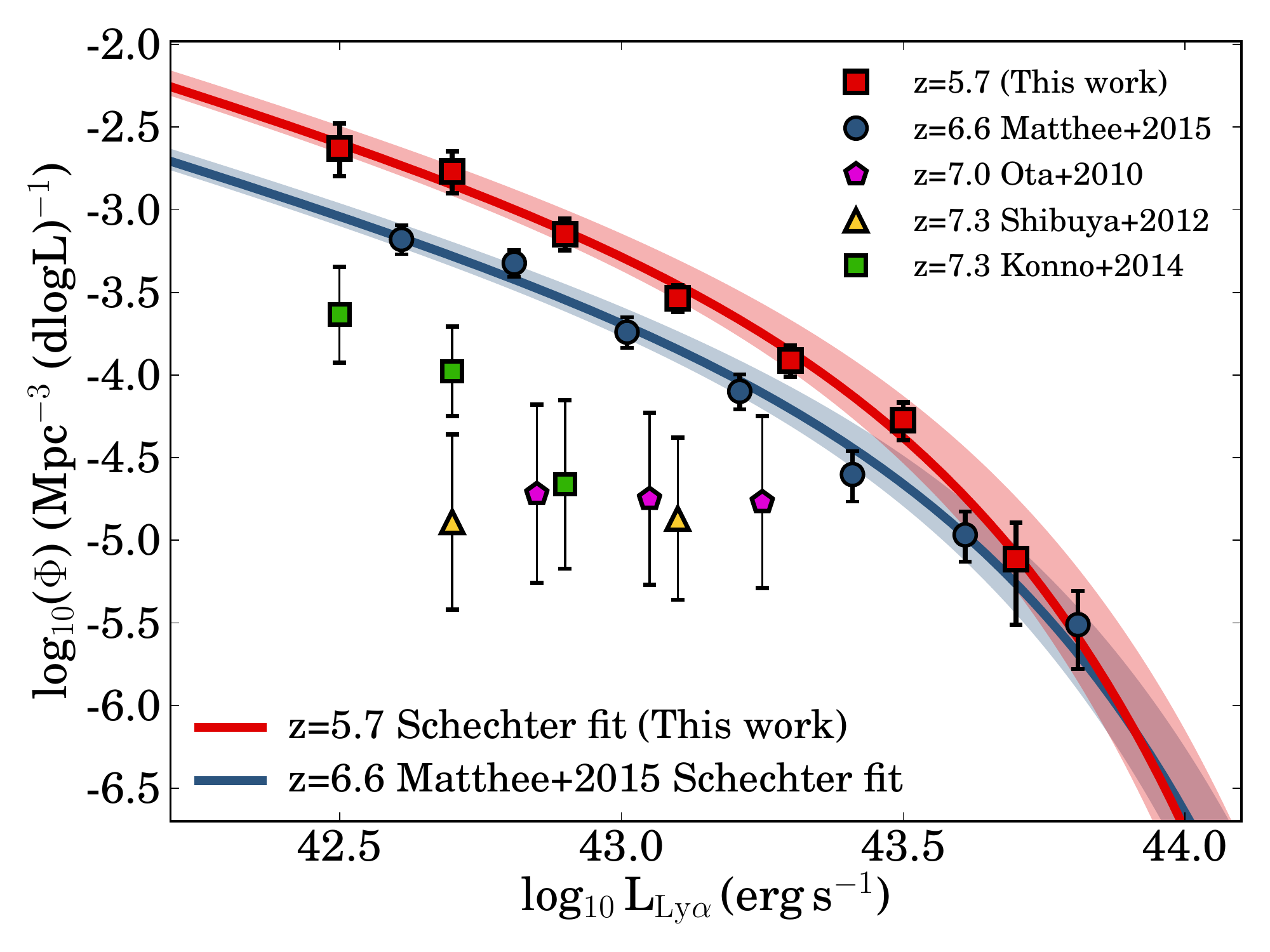}
\caption{Evolution of the Ly$\alpha$ LF from $z=5.7$ to $z=6.6$. The $z=6.6$ LF is our updated version from \citet{Matthee2015}, see Section \ref{recomp_z66}. The colored regions around the best Schechter fit show the 1$\sigma$ error in $L^*$. We observe a strong decrease in the number density of the fainter LAEs as we increase with redshift up to $z=6.6$ and also $z>7$ \citep{Ota2010,Shibuya2012,Konno2014}. This decrease can likely be explained by a more neutral IGM as we go deeper into the reionization epoch. However, there seems to be no evolution for the brighter sources, which can likely be explained by a preferential reionization around the brightest sources. There is currently a lack of comparable surveys at $z>7$ at the brightest luminosities.}
\label{fig:lfevo}
\end{figure*}

\subsubsection{Comparison with other $z=5.7$ surveys}

Several surveys have published LFs of $z=5.7$ LAEs, which we compare with our results (see Figure \ref{fig:lfindividual}). We compare our results with \cite{Westra2006}, \cite{Murayama2007} (COSMOS), \cite{Ouchi2008} (UDS) and \citet{Hu2010} (SA22-deep, SSA17, A370 and GOODS-N) in Figure \ref{fig:lfindividual}. While there are some differences between our selection criteria and the ones applied in these studies, overall we find very good agreement. Moreover, the variance that we see from field to field (see Figure \ref{fig:lfindividual}) is sufficient to explain any subtle differences between our results per field and those in the literature.

For the COSMOS field, \citet{Murayama2007} applies a much more conservative $\Sigma$ cut (corresponding to roughly $\Sigma>5$) which leads to missing fainter LAEs. The different $\Sigma$ cut, together with a different completeness correction (ours is based on line-flux or luminosity, while \citealt{Murayama2007} does a correction based on detection completeness) easily explains why our fainter luminosity bin (log$_{10}$L$_{\rm Ly\alpha}=42.9$\,erg\,s$^{-1}$) has a higher number density, which fully agrees with our UDS and SA22 estimates, along with those presented in \cite{Ouchi2008}.

Within the errors, our results are also fully consistent with those by \cite{Ouchi2008}, at all luminosities. Our brightest bin (log$_{10}$L$_{\rm Ly\alpha}=43.7$\,erg\,s$^{-1}$) is populated only by our COSMOS and SA22-wide fields, as those have the largest areas (sufficiently large to probe the bright end), but we note that the estimates from COSMOS and SA22-wide fully agree, while we are also in very good agreement with the results from \cite{Hu2010}. SA22-deep is both our smallest contiguous field and also the one with the highest number densities (although generally agreeing within the errors with the other fields, particularly given the variance seen). In the SA22-wide field we find number densities consistent with \citet{Ouchi2008} up to log$_{10}$L$_{\rm Ly\alpha}=43.5$ erg s$^{-1}$ and a brighter bin consistent with our COSMOS number density. The bright end of the Ly$\alpha$ LF seems to point towards a deviation from the Schechter fit presented in \citet{Ouchi2008}, better explained by a less accentuated exponential drop, or by a single power-law.

\subsubsection{The combined $z=5.7$ Ly$\alpha$ luminosity function}

We combine our data from the different fields to obtain a combined Ly$\alpha$ luminosity function at $z=5.7$. We show the results in Figure \ref{fig:lfevo} and Table \ref{tab:bins_combined}. 

We fit a Schechter function \citep{Schechter1976}, defined by three parameters: the power-law slope $\alpha$, the characteristic number density $\phi^\star$ and the characteristic luminosity $L^\star$.

In Table \ref{tab:schechterfit}, we present best-fit parameters of the Schechter function at $z=5.7$. We find the faint end slope $\alpha$ to be particularly steep: $\alpha=-2.3^{+0.4}_{-0.3}$. This is in very good agreement with recent results from \cite{Dressler2015} at the same redshift who found $\alpha$ to be $-2.35<\alpha<-1.95$ (while we find $-2.6<\alpha<-1.9$, 1\,$\sigma$). It is therefore clear that the Ly$\alpha$ luminosity function is very steep just after re-ionisation and may be steeper than the UV luminosity function at the same redshift \citep[$\alpha\approx-1.9$; e.g.][]{Bouwens2015}. Note that such a steep faint-end slope at $z=5.7$ is already preferred by the fit in \cite{Ouchi2008} and is consistent with theoretical expectations \citep{Gronke2015}.

We also fit our LF by fixing the faint-end slope to $\alpha=-2.0$ and $\alpha=-1.5$ and allow $\Phi^\star$ and $L^\star$ to vary. This allows our results to be directly compared with other studies which fixed $\alpha$ to the same values. The results are presented in Table \ref{tab:schechterfit}.

%
%
\begin{table}
\centering
\caption{Parameters for the best Schechter function fits for the Ly$\alpha$ LFs at $z=5.7$ and $z=6.6$ \citep[recomputed][]{Matthee2015}. We allow $\alpha$ to vary, but we also fix $\alpha$ to $-2.0$ and $-1.5$.}\label{tab:schechterfit}
\begin{tabular}{lccc}
\hline
{\bf Redshift} & $\bf \alpha$ & \bf log$_{10}$L$^\star_{{\rm Ly}\alpha}$ &  \bf log$_{10}\Phi^\star$\\
 & &  (erg\,s$^{-1}$) &  (Mpc$^{-3}$) \\
\hline
\boldmath$z=5.7$  & $-2.3^{+0.4}_{-0.3}$ & 43.42$^{+0.50}_{-0.22}$ & -4.02$^{+0.48}_{-0.93}$ \\
 & $-1.5$ (fix) & 43.06$^{+0.05}_{-0.04}$ & -3.25$^{+0.09}_{-0.10}$ \\
 & $-2.0$ (fix) & 43.25$^{+0.09}_{-0.06}$ & -3.63$^{+0.12}_{-0.16}$ \\
\hline
\boldmath$z=6.6$  & $-2.3^{+0.4}_{-0.3}$ & 43.45$^{+0.35}_{-0.18}$ & -4.48$^{+0.43}_{-0.68}$ \\
 & $-1.5$ (fix) & 43.12$^{+0.04}_{-0.03}$ & -3.73$^{+0.07}_{-0.06}$ \\
 & $-2.0$ (fix) & 43.30$^{+0.07}_{-0.05}$ & -4.13$^{+0.10}_{-0.10}$ \\
\hline
\end{tabular}
\end{table}

\subsection{Evolution from $\bf z=5.7$ to $\bf z\sim7$ and beyond} \label{sec:evolution}

In Section \ref{recomp_z66} we discuss the steps we took to obtain a comparable and updated $z=6.6$ Ly$\alpha$ luminosity function, based on \cite{Matthee2015}. We show the recomputed $z=6.6$ Ly$\alpha$ LF, and a comparison with our $z=5.7$ measurement in Figure \ref{fig:lfevo}. The recomputed $z=6.6$ LF is fully presented in Table \ref{tab:bins_combined}.

We find that both $z=6.6$ and $z=5.7$ are best fit with a very steep $\alpha$ of $\sim-2.3$. At a fixed $\alpha$, our results show a significant decline in the number density of the more ``typical"/faint Ly$\alpha$ emitters from $z=5.7$ to $z=6.6$, with $\phi^*$ declining by 0.5\,dex. However, and in very good agreement with \cite{Matthee2015}, we find little to no evolution at the bright end, with $L^*$ showing no significant evolution, or only a very weak increase of $\sim0.05-0.1$\,dex from $z=5.7-6.6$ (depending on $\alpha$). In practice, our results show that the number density of bright Ly$\alpha$ emitters (L$_{\rm Ly\alpha}>10^{43.5}$\,erg\,s$^{-1}$) shows no significant evolution from $z=5.7$ to $z=6.6$, confirming the results suggested in \cite{Matthee2015}. We note that while we discuss the luminosity functions in the context of their Schechter fits, the results presented hold if we fit them with e.g. single or double power-laws. At $z=6.6$, the spectroscopic confirmation of the sources responsible for these high Ly$\alpha$ luminosities is starting to reveal their uniqueness \citep[e.g. multi-component, very low metallicities, blue Ly$\alpha$ wings, range of sizes, see e.g. Himiko, MASOSA, CR7, COLA1;][]{Ouchi2013,Sobral2015,Hu2016}, providing important hints that may explain how these sources have been able to likely reionize their surroundings already at $z\sim7$. Further observations will be able to confirm a larger, statistical sample at $z\sim7$, but our new sample at $z=5.7$ is uniquely suited to be directly compared.

Figure \ref{fig:lfevo} also presents results from several $z>7$ narrow band surveys from the literature, which we compare with $z=6.6$ and $z=5.7$. The trend that we see from $z=5.7$ to $z=6.6$ of significant decrease in the number density of faint Ly$\alpha$ emitters seems to continue at a fast pace to $z\sim7$ and beyond \citep{Ota2010,Shibuya2012,Konno2014}. We provide a more detailed discussion about the differential evolution of the Ly$\alpha$ as an imprint of reionization in Section \ref{sec:discussion}. There is currently a lack of comparable surveys at $z>7$ at the brightest luminosities, so it is not yet possible to test whether the lack of evolution at the bright end still holds at $z>7$.

\subsection{The Ly$\alpha$ sizes and evolution at $\bf z=5.7-6.6$} \label{sec:aperture}

Since the Ly$\alpha$ transition is resonant, Ly$\alpha$ photons scatter in a medium with neutral hydrogen. Because of this, Ly$\alpha$ photons tend to escape over much large radii than their UV and H$\alpha$ counterparts, making them observable as Ly$\alpha$ haloes \citep[e.g.][]{Rauch2008,Steidel2011,Momose2014,Matthee2016}. Therefore, the aperture that is used to measure Ly$\alpha$ is critical \citep[e.g.][]{Wisotzki2016}. Typically, LAE surveys have attempted to take extended Ly$\alpha$ emission into account by using {\sc mag-auto} measurements \citep[e.g.][]{Ouchi2010,Konno2016} or relatively large apertures \citep[e.g.][who use 3$''$ apertures at $z=5.7$]{Murayama2007,Hu2010}. However, the total measured magnitude with {\sc mag-auto} depends on the depth of the narrow-band imaging, such that a comparison between surveys and redshifts is challenging, particularly as \citet{Wisotzki2016} show that Ly$\alpha$ extends well beyond the typical limiting surface brightness of narrow-band surveys.

%
%
\begin{figure}
\includegraphics[width=8.8cm]{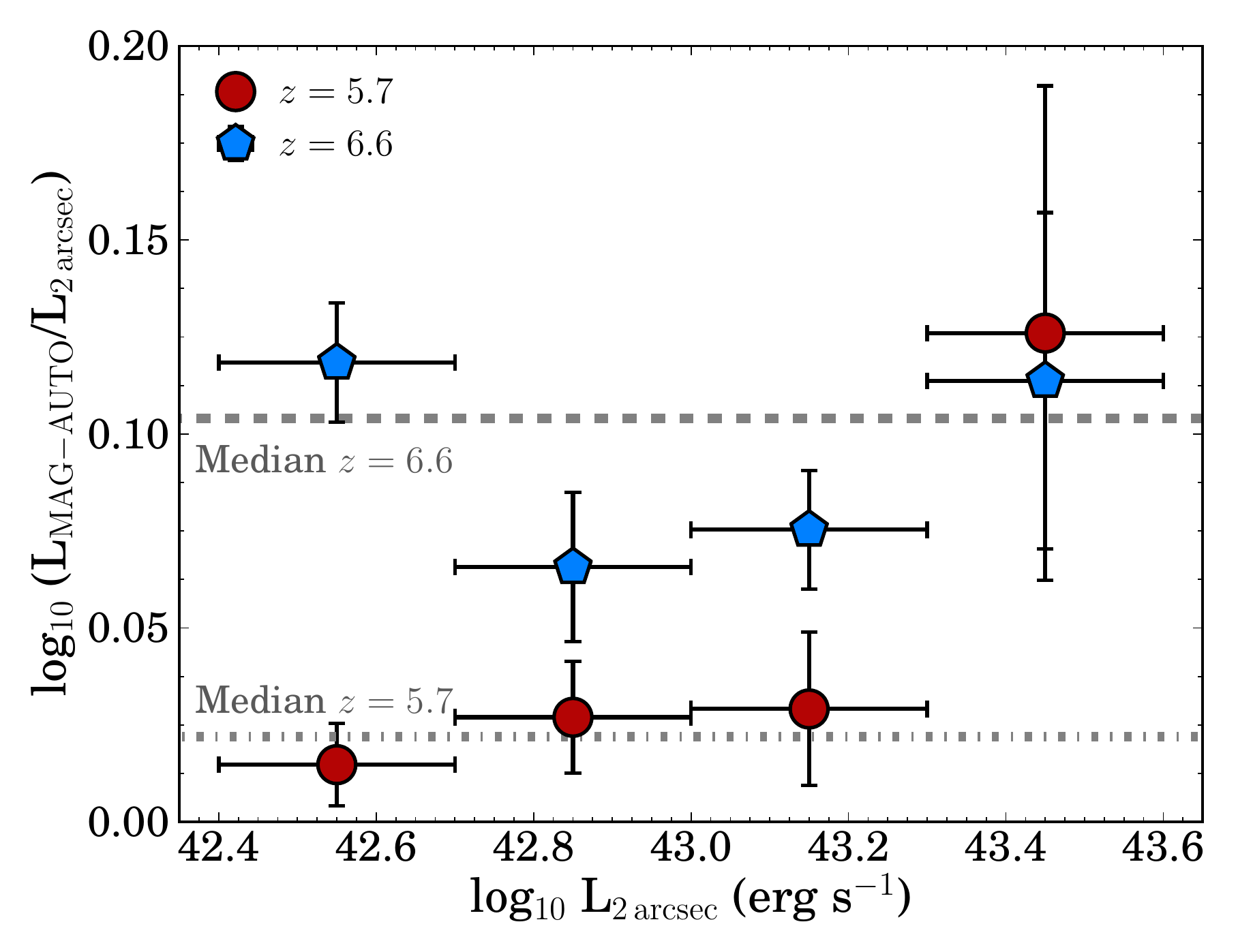}
\caption{The median difference in {\sc mag-auto} luminosity and luminosity within 2$''$ apertures in bins of the 2$''$ aperture Ly$\alpha$ luminosity for LAE samples at $z=5.7$ and $z=6.6$. The dashed and dashed-dotted grey lines indicate the median of all LAEs in the sample, which is obviously dominated by low luminosity sources. At both redshifts, more centrally luminous LAEs also have relatively more flux at larger radii (which is captured by {\sc mag-auto}). At faint central luminosities LAEs at $z=6.6$ appear more extended, which could be due to increased scattering in HI around galaxies. We note that this may be one of the causes for the apparent evolution in the Ly$\alpha$ LF, and may also be important to consider when interpreting the spectroscopic follow-up of UV-selected galaxies with low Ly$\alpha$ luminosities, as slits will recover even less of the total flux.}
\label{fig:magauto_magaper} 
\end{figure}

While we use fixed 2$''$ apertures in similar excellent seeing conditions at both $z=5.7$ and $z=6.6$ (as this allows to understand the completeness and selection function in an optimal way; c.f. \citealt{Matthee2015}), we correct for any flux missed as described in Section \ref{aper_correction}.

\cite{Matthee2015} found that 2$''$ apertures systematically underestimate Ly$\alpha$ luminosities at $z=6.6$ (compared to the {\sc mag-auto}) with a median offset of 0.11 dex over the spectroscopically confirmed sample of LAEs (confirmed in \citealt{Ouchi2010}). Here we extend this analysis to the full sample of sources at both $z=5.7$ and $z=6.6$. We find that the median offset between the {\sc mag-auto} luminosity and the 2$''$ aperture offset at $z=6.6$ is 0.11\,dex, while it is only 0.02 dex at $z=5.7$; see Figure \ref{fig:magauto_magaper}. The latter explains why our 2$''$ measurements result in very similar number densities as literature studies with larger apertures at $z=5.7$, see Fig. \ref{fig:lfindividual}.

By splitting the sample of LAEs in bins of Ly$\alpha$ luminosity (in 2$''$ apertures), we find that at $z=5.7$ the offset increases slightly with increasing Ly$\alpha$ luminosity (see Fig. $\ref{fig:magauto_magaper}$). Specifically, the most luminous LAEs have larger Ly$\alpha$ haloes (and more flux at larger radii) than the typical fainter ones. Interestingly, we find a different behaviour at $z=6.6$. While the brightest $z=6.6$ Ly$\alpha$ seem to be as extended as those at $z=5.7$ (these are the ones that may have already been able to fully ionise the surrounding environment), fainter Ly$\alpha$ emitters at $z=6.6$ are all more extended than comparable sources at $z=5.7$. Together with the differential evolution of the Ly$\alpha$ LF, our results provide strong evidence for reionization effects being much stronger for the faint sources than for the bright ones. We discuss this trend further in Section \ref{sec:discussion}. 

A similar but more careful analysis of the extent of Ly$\alpha$ emission at $z=5.7-6.6$ than our own has been done by \cite{Momose2014}, who created stacked narrow band and broad band images of the LAEs in UDS from \citet{Ouchi2008,Ouchi2010}. They observed that Ly$\alpha$ is extended, being more extended than their UV counterpart (while also being more extended than the PSF of their images; a similar trend is found for individual LAEs by e.g. \citealt{Wisotzki2016}). \citealt{Momose2014} found evidence of an increase in the scale length of Ly$\alpha$ from $z=5.7$ to $z=6.6$. However, they did not separate their sample in bins of luminosity and their results are obtained with median stacking. This means that the faintest sources dominate (as there are more faint sources than luminous ones) and that these results are more representative of a ``typical" LAE, with L$_{\rm Ly\alpha}\sim10^{42.6}$\,erg\,s$^{-1}$. The median evolution in the scale length of Ly$\alpha$ haloes from LAEs estimated in \citet{Momose2014} is thus consistent with the difference between {\sc mag-auto} and 2$''$ measurements that we find for relatively faint LAEs between $z=5.7$ and $z=6.6$.

\section{Discussion: Imprints from reionization?} \label{sec:discussion}

As noted before, the observed Ly$\alpha$ luminosity at a fixed spatial scale is expected to decrease in the reionization era, as an increasingly neutral IGM scatters Ly$\alpha$ photons into larger, extended haloes \citep[e.g.][]{Dijkstra2014}. Our results are consistent with witnessing such predictions directly. Here we discuss the differences we observe in the Ly$\alpha$ luminosity function between $z=5.7$ and $z=6.6$, and also our results on the extent of Ly$\alpha$ emitters at $z=5.7$ and $z=6.6$. For earlier work, see e.g. \cite{Dijkstra2007}, \citet{Ouchi2010} and \citet{Hu2010}.

We observe strong differential evolution of the Ly$\alpha$ LF from $z\sim6$ to $z\sim7$, with a significant decrease ($-0.5$\,dex) in the number density for Ly$\alpha$ luminosities below $L^*$. The drop in the observability of faint LAEs may well be explained by a larger fraction of neutral IGM at $z>6$ caused by reionization not being completed. The brightest emitters would not suffer from such a decline because their strong Ly$\alpha$ emission is easier to be observed, as previously illustrated by the simple toy model in \cite{Matthee2015}. This model assumes that the Ly$\alpha$ luminosity scales with the ionising output and LAEs are only observed if they are either capable of ionising the IGM around them, or are strongly clustered. To first order, a stronger ionising output for brighter LAEs is expected because Ly$\alpha$ is a recombination line (such that at fixed escape fraction, a higher Ly$\alpha$ luminosity scales with the number of ionising photons). Also, as shown in \cite{MattheeGALEX2016}, LAEs at $z=2.2$ typically produce more ionising photons per unit UV luminosity than more typical galaxies such as H$\alpha$ emitters (HAEs). Furthermore, as hypothesised by \cite{DijkstraGronke2016}, ISM conditions which favor the escape of Ly$\alpha$ photons also likely favor the escape of Lyman continuum (LyC) photons (for example due to a porous ISM), such that in addition to producing more ionising photons, LAEs could also leak more ionising photons into the IGM.

Recent evidence from \citet{Stark2016} shows that the fraction of bright UV selected galaxies (LBGs) with strong Ly$\alpha$ emission is much higher than was previously found \citep[e.g.][]{Schenker2014,Pentericci2014,Schmidt2016} when they are selected on strong nebular lines (e.g. H$\beta$/[O{\sc iii}]). This is likely because UV-bright galaxies are in over-dense regions and emit copious amount of ionising radiation (inferred from observed high ionization UV lines as C{\sc iii]} and their high EW optical nebular lines). Such conditions may also favor the production of Ly$\alpha$ photons and lead to larger ionised bubbles. Therefore, these observations are in principle consistent with the observed evolution of the Ly$\alpha$ LF, where we observe reionization completing first around luminous LAEs. 

A unique benefit of narrow-band Ly$\alpha$ observations over (slit) spectroscopy is that narrow-band imaging gives information on the spatial extent of Ly$\alpha$ emission, which could be connected to the neutral fraction of the IGM \citep[e.g.][]{DijkstraLoeb2008}. As we show in Figure $\ref{fig:magauto_magaper}$, we find that the median difference between 2$''$ apertures and the total magnitude (as observed with {\sc mag-auto}) is much smaller at $z=5.7$ than at $z=6.6$. Most interestingly, the major difference is found at the faintest luminosities. At $z=6.6$, LAEs which have a low central luminosity have a relatively much larger total luminosity than at $z=5.7$. This means that at a fixed surface brightness limit (note that the limiting surface brightness at $z=6.6$ is actually even slightly higher), faint LAEs are more extended at $z=6.6$ than at $z=5.7$. For more luminous LAEs the difference is much smaller. This effect can easily be explained in the framework of the \cite{Matthee2015} toy-model: faint LAEs are surrounded by a relatively more neutral IGM, such that there is more resonant scattering leading to more extended emission. 

The evolution of the Ly$\alpha$ LF and the extent of Ly$\alpha$ for different luminosities may very well be explained by a patchy reionization scenario where the IGM is ionised first around luminous LAEs. However, internal effects from galaxies may also be important. Furthermore, studying the clustering of both bright and faint LAEs and how it evolves from e.g. $z=5.7$ to $z=6.6$ and beyond \citep[e.g.][]{Mesinger2010,Ouchi2010} will provide the extra, necessary constraints. A similar analysis with future larger samples of LAEs (for example from the Hyper Suprime Cam survey) will be very useful to confirm the observed trends.

Our results also mean that a careful approach is required in order to interpret the observed Ly$\alpha$ fraction for samples of LBGs at different redshifts in terms of a varying neutral fraction due to reionization, because different samples of LBGs show very different Ly$\alpha$ fractions. \citet{CurtisLake2012} found a remarkably high fraction of strong LAEs amongst luminous LBGs, \citet{Stark2016} found a higher Ly$\alpha$ fraction for LBGs selected on strong nebular emission and \citet{Erb2016} found that $z\sim2$ galaxies with extreme line ratios have high Ly$\alpha$ fractions. Moreover, our results show that typical, faint Ly$\alpha$ emitters become more extended as we go into the reionization epoch, with the same (or even less) flux being spread over larger areas. This is an additional challenge for the traditional slit spectroscopy follow-up, which will struggle to detect any Ly$\alpha$ if the flux is significantly extended.

\section{Conclusions} \label{sec:conclusions}

We have constructed the largest Ly$\alpha$ narrow band survey at $z=5.7$, when re-ionization is close to complete. We have surveyed a total area of $7$\,deg$^2$ and a volume of $6.3\times10^6$Mpc$^3$ at $z=5.7$, covering the COSMOS, UDS and SA22 fields. Here we summarize the main conclusions:

\begin{itemize}

\item By identifying strong line-emitters with a Lyman break, we find 514 LAE candidates at $z=5.7$ with EW$_{0} > 25 $ {\AA} (EW$_0\sim25-1000$\,\AA) and luminosities ranging from $10^{42.5}-10^{44}$ erg s$^{-1}$, in a single, homogeneous data-set.

\item We find that cosmic variance plays a major role, with variations of $\pm0.4$\,dex in number densities of Ly$\alpha$ emitters from field to field.

\item By combining all our fields and overcoming cosmic variance, we find that the faint end slope of the $z=5.7$ Ly$\alpha$ luminosity function is very steep, with $\alpha=-2.3^{+0.4}_{-0.3}$. If we fix $\alpha=-2.0$, we find $L^\star=10^{43.22^{+0.08}_{-0.05}}$\,erg\,s$^{-1}$ and $\Phi^\star=-3.60^{+0.12}_{-0.16}$\,Mpc$^{-3}$.

\item We also present an updated $z=6.6$ Ly$\alpha$ luminosity function, based on comparable volumes, and obtained with the same methods, which we directly compare with that at $z=5.7$.

\item We find significant evolution from $z=5.7$ (after re-ionization) to $z=6.6$ (within the epoch of re-ionization) at the faint end. We find that the fainter the luminosity, the stronger the drop in the number density of Ly$\alpha$ emitters. The strong decrease of the number density of faint Ly$\alpha$ emitters continues to $z\sim7$.

\item At bright Ly$\alpha$ luminosities (L$_{\rm Ly\alpha}>10^{43.5}$\,erg\,s$^{-1}$) we find no evolution in the number density of Ly$\alpha$ emitters when we enter the re-ionization era. This is consistent with bright Ly$\alpha$ emitters being preferentially observable because they already are in ionized bubbles even at $z\sim7$.

\item Faint Ly$\alpha$ emitters at $z=6.6$ show more extended haloes than those at $z=5.7$, suggesting that neutral Hydrogen plays a more important role of scattering Ly$\alpha$ photons at $z=6.6$.

\end{itemize}

All together, our results indicate that we are observing patchy reionization happening first around the brightest Ly$\alpha$ emitters, allowing the number densities of those sources to remain unaffected by the increase of neutral Hydrogen from $z\sim5$ to $z\sim7$. We observe a preferential evolution of the faint end of the Ly$\alpha$ LF from $z=5.7$ to $z=6.6$. There is a decrease in the faint end while the bright end shows little to no evolution. We also observe no evolution in the sizes of the brighter emitters, which could be interpreted as showing no evidence of extra scattering around them from $z=5.7$ to $z=6.6$, while faint sources show a significant difference, presenting much more flux at larger radii, which could be explained by faint LAEs being located in a more neutral IGM leading to more resonant scattering and extended emission. The spectroscopic confirmation of relatively bright Ly$\alpha$ emitters beyond $z\sim7$ and approaching $z\sim9$ \citep[][]{Oesch2015,Zitrin2015} may already be hinting that our results may hold to even higher redshifts.

The nature and diversity of bright Ly$\alpha$ sources at $z=6.6$, which we find to have essentially the same number density as those at $z=5.7$, are starting to be unveiled. Spectroscopic follow up \citep[e.g.][]{Ouchi2013,Sobral2015,Zabl2015,Hu2016}, detailed modelling \citep[e.g.][]{Hartwig2015,DijkstraG2016,Agarwal2016,Visbal2016,Smidt2016,Smith2016} and other observations with HST and ALMA \citep[][]{Ouchi2013,Sobral2015,Schaerer2015,Bowler2016} are revealing a surprising diversity. Current results indicate that these sources may have a range of powering sources (from metal poor populations to multiple stellar populations and also AGN, including potentially direct collapse black holes). Regardless of their nature, their observability requires the production and emission of the necessary amount of ionising LyC photons capable of ionising a large enough local bubble to make them observable as bright Ly$\alpha$ sources already at $z=6.6$. Thus, even though these sources are not as abundant as the more typical, faint sources, they may well play an important role in cosmic reionization, at least at very early stages, a scenario which would be in agreement with what is seen by \cite{MattheeGALEX2016}. Further observations of our sample of bright $z=5.7$ sources and of much larger, statistical samples at $z\sim5-7$ will certainly shed light over many of the current open questions, while the availability of JWST will provide a revolutionary window into the physical conditions within these sources.

\section*{Acknowledgements}

We thank the anonymous referee for useful and constructive comments and suggestions which greatly improved the quality and clarity of our work. The authors acknowledge financial support from the Netherlands Organisation for Scientific research (NWO) through a Veni fellowship. SS and DS acknowledge funding from FCT through a FCT Investigator Starting Grant and Start-up Grant (IF/01154/2012/CP0189/CT0010). SS also acknowledges support from FCT through the research grants UID/FIS/04434/2013 and PTDC/FIS-AST/2194/2012. JM acknowledges a Huygens PhD fellowship from Leiden University.

Based on observations with the Subaru Telescope (Program IDs: S05B-027, S06A-025, S06B-010, S07A-013, S07B-008, S08B-008, S09A-017, S14A-086). Based on observations made with ESO Telescopes at the La Silla Paranal Observatory under programme ID 294.A-5018. Based on observations obtained with MegaPrime/Megacam, a joint project of CFHT and CEA/IRFU, at the Canada-France-Hawaii Telescope (CFHT) which is operated by the National Research Council (NRC) of Canada, the Institut National des Science de l'Univers of the Centre National de la Recherche Scientifique (CNRS) of France, and the University of Hawaii. This work is based in part on data products produced at Terapix available at the Canadian Astronomy Data Centre as part of the Canada-France-Hawaii Telescope Legacy Survey, a collaborative project of NRC and CNRS. Based on data products from observations made with ESO Telescopes at the La Silla Paranal Observatory under ESO programme ID 179.A-2005 and on data products produced by TERAPIX and the Cambridge Astronomy Survey Unit on behalf of the UltraVISTA consortium. We are grateful to the CFHTLS, COSMOS-UltraVISTA, UKIDSS, SXDF and COSMOS survey teams. Without these legacy surveys, this research would have been impossible.

The authors wish to recognize and acknowledge the very significant cultural role and reverence that the summit of Mauna Kea has always had within the indigenous Hawaiian community. We are most fortunate to have the opportunity to conduct and explore observations from this mountain.

Finally, the authors acknowledge the unique value of the publicly available programming language {\sc Python}, including the {\sc numpy}, {\sc pyfits}, {\sc matplotlib}, {\sc scipy} and {\sc astropy} \citep{Astropy2013} packages.

\bibliographystyle{mnras}
\bibliography{myBib}


\appendix

\section{Filter profile corrections and LFs} \label{corr_proff}

\begin{figure}
  \centering
  \includegraphics[width=0.5\textwidth]{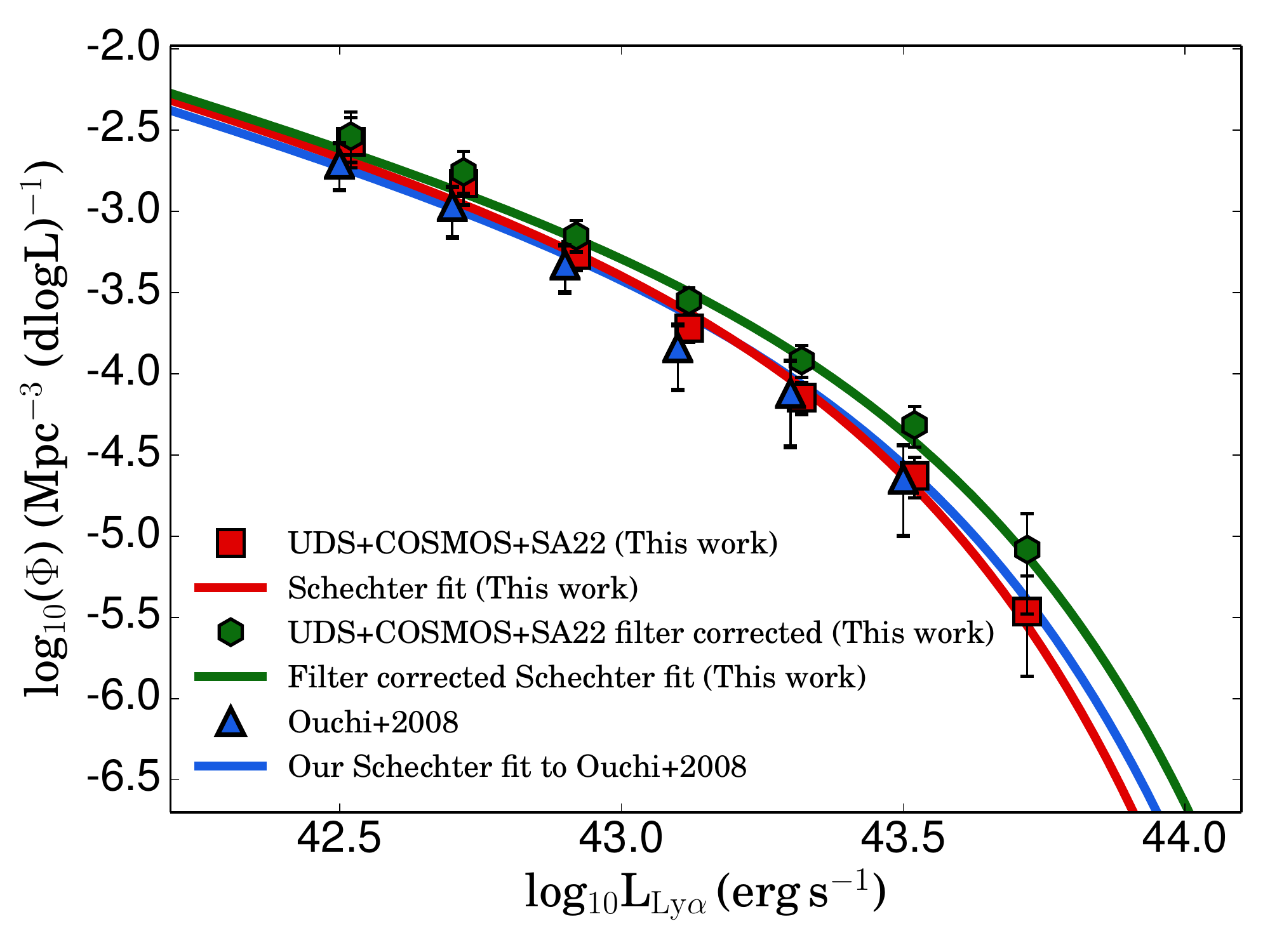}
  \caption{The number densities in luminosity bins from our survey in the UDS, COSMOS and SA22 fields (red squares) and the bins from \citet{Ouchi2008} in blue triangles. A small luminosity correction of +0.02 was applied to our luminosity bins to correct for extended emission (this correction is discussed in \S\ref{sec:aperture}). The Schechter fits to the luminosity bins from our study agrees very well with \citet{Ouchi2008}. In green we also show the luminosity bins from this work after we apply a filter profile bias correction (we estimate this correction in \S\ref{sec:filtercor}) and the corrected LF Schechter fit. The effect of this correction is strongest at the brightest bins.} 
  \label{fig:filtercorrect}
\end{figure}

Figure \ref{fig:filtercorrect} shows the effect of our filter profile corrections. We show the completeness corrected number densities of LAEs in bins of Ly$\alpha$ luminosity for individual fields at $z=5.7$ (Table \ref{tab:bins}) and for the combined coverage at $z=5.7$ and $z=6.6$ (Table \ref{tab:bins_combined}). 

%
%

%
%

\begin{table*}
\caption{For each field we present the median line-flux completeness per bin, which we use to correct the $z=5.7$ number densities. We only consider number densities from sub-fields with a line-flux completeness higher than 25\%.}
\begin{tabular}{@{}ccccccc@{}}
\hline
\bf Luminosity bins & \multicolumn{4}{c}{\bf Line-flux completeness} \\
log$_{10}$L [erg s$^{-1}$] & \multicolumn{4}{c}{Percentage [\%]} \\
& (UDS) & (COSMOS) & (SA22-deep) & (SA22-wide)\\
\hline
$42.5\pm0.1$ & 27 &$ <25$ &$ <25$ &$ <25$\\
$42.7\pm0.1$ & 30 &$ <25$ &$ <25$ &$ <25$\\
$42.9\pm0.1$ & 45 & 37 & 36 & $ <25$\\
$43.1\pm0.1$ & 53 & 54 & 56 & $ <25$\\
$43.3\pm0.1$ & 61 & 65 & 68 & 51\\
$43.5\pm0.1$ & 73 & 73 & 77 & 63\\
$43.7\pm0.1$ & 83 & 80 & 84 & 74\\
\hline
\end{tabular}
\label{tab:completeness}
\end{table*}

%
%
\begin{table}
\caption{The completeness corrected number density of LAEs in the different surveyed fields at $z=5.7$.}
\begin{tabular}{@{}ccccccc@{}}
\hline
\bf Field & \bf Luminosity bin & \bf Number density\\
& log$_{10}$L [erg s$^{-1}$] & log$_{10}\Phi$/dlogL [Mpc$^{-3}$] \\
\hline
\bf UDS&$42.5\pm0.1$ & -2.57$_{-0.16}^{+0.15}$\\
  &$42.7\pm0.1$ & -2.82$_{-0.13}^{+0.13}$\\
  &$42.9\pm0.1$ & -3.37$_{-0.15}^{+0.13}$\\
  &$43.1\pm0.1$ & -3.94$_{-0.20}^{+0.16}$\\
  &$43.3\pm0.1$ & -4.37$_{-0.33}^{+0.21}$\\
\hline
\bf COSMOS & $42.9\pm0.1$ & -3.30$_{-0.13}^{+0.12}$\\
  &$43.1\pm0.1$ & -3.81$_{-0.13}^{+0.11}$\\
  &$43.3\pm0.1$ & -4.40$_{-0.21}^{+0.16}$\\
  & $43.5\pm0.1$ & -4.93$_{-0.39}^{+0.22}$\\
  & $43.7\pm0.1$ & -5.42$_{-\infty}^{+0.32}$\\
\hline
\bf SA22-deep&$42.9\pm0.1$ & -3.09$_{-0.12}^{+0.11}$\\
  &$43.1\pm0.1$ & -3.37$_{-0.11}^{+0.09}$\\
  &$43.3\pm0.1$ & -3.84$_{-0.18}^{+0.14}$\\
  &$43.5\pm0.1$ & -4.50$_{-0.38}^{+0.21}$\\
\hline
\bf SA22-wide&$43.3\pm0.1$ & -4.07$_{-0.13}^{+0.11}$\\
  &$43.5\pm0.1$ & -4.41$_{-0.16}^{+0.13}$\\
  &$43.7\pm0.1$ & -5.33$_{-0.56}^{+0.26}$\\
\hline
\end{tabular}
\label{tab:bins}
\end{table}

%
%
\begin{table*}
\caption{The completeness and filter profile bias corrected luminosity functions at $z=5.7$ and $z=6.6$ from this study. Note that we corrected the bins for extended emission (see Section \ref{sec:aperture}).}
\begin{tabular}{@{}ccccccc@{}}
\hline
\bf Redshift & \bf Luminosity bin & \bf Volume & \bf Observed number density & \bf Corrected number density\\
 & log$_{10}$L [erg s$^{-1}$] & [$10^{6}$ Mpc$^3$] & log$_{10}\Phi$/dlogL [Mpc$^{-3}$] & log$_{10}\Phi$/dlogL [Mpc$^{-3}$]\\
\hline
\boldmath $z=5.7$ &$42.52\pm0.1$ & 0.19 & -3.16$_{-0.09}^{+0.08}$ & -2.63$_{-0.17}^{+0.16}$\\
  &$42.72\pm0.1$ & 0.65 & -3.32$_{-0.06}^{+0.05}$ & -2.77$_{-0.13}^{+0.12}$\\
  &$42.92\pm0.1$ & 3.09 &-3.65$_{-0.04}^{+0.04}$ & -3.15$_{-0.10}^{+0.10}$\\
  &$43.12\pm0.1$ & 3.09 &-3.89$_{-0.05}^{+0.05}$ & -3.54$_{-0.08}^{+0.08}$\\
  &$43.32\pm0.1$ & 6.30 & -4.34$_{-0.06}^{+0.05}$ & -3.91$_{-0.10}^{+0.09}$\\
  &$43.52\pm0.1$ & 6.30 & -4.70$_{-0.10}^{+0.08}$ & -4.27$_{-0.12}^{+0.11}$\\
  &$43.72\pm0.1$ & 6.30 &-5.62$_{-0.37}^{+0.20}$ & -5.12$_{-0.40}^{+0.22}$\\
\hline
\boldmath $z=6.6$ & $42.61\pm0.1$ & 0.38 & -3.46$_{-0.08}^{+0.09}$ &  -3.18$_{-0.09}^{+0.08}$\\
  &$42.81\pm0.1$ & 0.64 & -3.59$_{-0.07}^{+0.08}$ & -3.32$_{-0.08}^{+0.08}$\\
  &$43.01\pm0.1$ & 1.07 & -4.01$_{-0.09}^{+0.11}$ & -3.74$_{-0.10}^{+0.09}$\\
  &$43.21\pm0.1$ & 1.73 & -4.42$_{-0.11}^{+0.14}$ &  -4.10$_{-0.11}^{+0.10}$\\
  &$43.41\pm0.1$ & 1.73 & -4.94$_{-0.18}^{+0.30}$ & -4.60$_{-0.16}^{+0.14}$\\
  &$43.61\pm0.1$ & 4.18 & -5.34$_{-0.18}^{+0.31}$ & -4.97$_{-0.16}^{+0.14}$\\
  &$43.81\pm0.1$ & 4.18 & -5.97$_{-0.26}^{+0.31}$ & -5.51$_{-0.26}^{+0.20}$\\
\hline
\end{tabular}
\label{tab:bins_combined}
\end{table*}


\bsp	
\label{lastpage}
\end{document}